\documentclass[12pt]{ecoc02}
\usepackage{amsmath,amssymb,epsf}

\usepackage{mathptmx,graphicx}
\usepackage{algorithm}% http://ctan.org/pkg/algorithms
\usepackage{algpseudocode}
\usepackage{pdfpages,cite}

\begin{document}

\title{Piecewise Constant Sequential Importance Sampling for Fast Particle Filtering}

\markboth{\"Omer Demirel, Ihor Smal, Wiro J.~Niessen, Erik Meijering and Ivo F.~Sbalzarini}{Approximate Sequential Importance Sampling for Fast Particle Filtering}

\author{\"Omer Demirel$^{*}$, Ihor Smal$^{\dagger}$, Wiro J.~Niessen$^{\dagger}$, Erik Meijering$^{\dagger}$ and Ivo F.~Sbalzarini$^{*}$}

\address{$^{*}$\textit{MOSAIC Group, Center of Systems Biology Dresden (CSBD),
Max Planck Institute of Molecular Cell Biology and Genetics,
Pfotenhauerstr. 108, 01307 Dresden, Germany. email: \{demirel,ivos\}@mpi-cbg.de}\\
$^{\dagger}$\textit{Biomedical Imaging Group Rotterdam, Departments of Medical Informatics and Radiology,  
Erasmus MC -- University Medical Center Rotterdam, Rotterdam, The Netherlands. email: \{i.smal,w.niessen\}@erasmusmc.nl, meijering@imagescience.org}}


\begin{abstract}
Particle filters are key algorithms for object tracking under non-linear, non-Gaussian dynamics. The high computational cost of particle filters, however, hampers their applicability in cases where the likelihood model is costly to evaluate, or where large numbers of particles are required to represent the posterior. We introduce the piecewise constant sequential importance sampling/resampling (pcSIR) algorithm, which aims at reducing the cost of traditional particle filters by approximating the likelihood with a mixture of uniform distributions over pre-defined cells or bins. The particles in each bin are represented by a dummy particle at the center of mass of the original particle distribution and with a state vector that is the average of the states of all particles in the same bin. The likelihood is only evaluated for the dummy particles, and the resulting weight is identically assigned to all particles in the bin. We derive upper bounds on the approximation error of the so-obtained piecewise constant function representation, and analyze how bin size affects tracking accuracy and runtime. Further, we show numerically that the pcSIR approximation error converges to that of sequential importance sampling/resampling (SIR) as the bin size is decreased. We present a set of numerical experiments from the field of biological image processing and tracking that demonstrate pcSIR's capabilities. Overall, we consider pcSIR a promising candidate for simple, fast particle filtering in generic applications, especially in those with a costly likelihood update step.\vspace{1pc}
\end{abstract}

\twocolumn

\maketitle

\section{Introduction}

Since their inception, sequential Monte Carlo (SMC) resampling methods (a.k.a., particle filters)~\cite{Doucet2000,Doucet2001} have emerged as a useful tool to estimate and track targets with non-linear and/or non-Gaussian dynamics. Unlike the Kalman filter~\cite{Kalman1960} and its variants~\cite{Wan2000}, particle filters (PF) do not use a fixed functional form of the posterior probability density function (PDF). Instead, they employ a finite number of points, called ``particles'', to discretely approximate the posterior probability density function (PDF) in state space~\cite{Thrun2005}.

A standard PF algorithm consists of two parts: (i) sequential importance sampling (SIS) and (ii) resampling~\cite{Doucet2001}. A popular combined implementation of these two parts is the sequential importance resampling (SIR) algorithm. Depending on the application, SIR may need a large number of particles to adequately sample the state space. This demands substantial computational resources that scale linearly with the number of particles and may hinder actualization of many practical real-time applications.  

Here, we introduce the \textit{piecewise constant SIR} (pcSIR) algorithm, which reduces the computational cost of SIR while providing tracking accuracy comparable to standard SIR. The main idea behind pcSIR is to group particles in state space (i.e., creating \textit{bins}) and to represent each group of particles by a single \textit{representative particle}. Only the weight of this representative dummy particle is then updated. We choose the dummy particle to sit in the center of mass of the group of particles it represents and to carry the mean properties of all the particles in the respective group. This is inspired by first-order multipole expansions from particle function approximation theory~\cite{greengard1987fast}. Once the weight of the dummy particle is computed, all other particles in the same group receive the same weight, which is copied from the dummy instead of being re-computed through the likelihood model for each individual particle, as in the original SIR. This way, an pcSIR-based PF can outperform a classical SIR-based PF by orders of magnitude in overall runtime in applications where evaluation of the likelihood is computationally expensive. Expensive likelihoods are particularly common when tracking objects in images, where each likelihood evaluation entails a numerical simulation of the image-formation process (see, e.g., Ref.~\cite{smaltmi}).

We outline the mathematical roots of pcSIR and derive an upper bound on the expected approximation error with respect to the chosen \textit{bin} (i.e., Cartesian mesh \textit{cell} in 2D) size. This error stems from the point-wise approximation of the likelihood function and is quantified using mid-point Riemann-sum error analysis~\cite{Davis1967,Thomas1984}. We numerically quantify the errors in the state estimates (based on the posterior distribution) obtained by SIR and pcSIR as a function of the number of particles used, and show that there is almost no difference between SIR and pcSIR in terms of tracking accuracy. Furthermore, with a focus on biological image processing, we show that relating the bin size to the pixel size of an image provides satisfactory, and sometimes even higher-quality results in pcSIR compared with standard SIR.

The structure of this manuscript is as follows: Section~2 summarizes similar approaches to PF for state estimation. Section~3 recapitulates the classical SIR algorithm, whereas Section~4 introduces our new pcSIR method, discusses the theoretical framework behind pcSIR, and provides detailed pseudocode. In Section~5, we benchmark pcSIR against SIR in terms of tracking accuracy, runtime, and error convergence using two different likelihood functions and different types of images. Finally, Section~6 discusses the  results and concludes the manuscript with an outlook. 

\section{Related Works}
\label{sec:relwork}
Recent years have seen great interest in challenging tracking problems where the targets usually have non-linear and/or non-Gaussian dynamics. Two nonparametric algorithms, namely the histogram filters (HF) and particle filters (PF), stand out amongst others as main classes of algorithms that successfully tackle difficult tracking problems~\cite{Thrun2005}. In both variants, posterior distributions are approximated by a finite set of values.  

In HF, the state space is decomposed into smaller -- usually rectangular -- boxes and only a single value is used to represent the cumulative posterior in each box. In a mathematical sense, HFs can be seen as piecewise constant approximations to a posterior distribution. The size and the number of the boxes affect the computational runtime and tracking accuracy of an application. 

In PF, random samples (i.e., point particles) are drawn from the posterior distribution and typically a large number of particles is required to track targets successfully. This increases the computational resources needed, and many PF-based applications are limited by their computational cost.

Combining ideas from PF and HF, the box particle filter (BPF)~\cite{Abdallah2008} uses box-shaped particles. While BPF resembles HF with mobile boxes, these box particles are propagated based on interval analysis~\cite{Jaulin2001}, which is fundamentally different from PF and HF. BPF is especially useful in situations where imprecise measurements yield wide posterior densities~\cite{Gning2013}. Despite its advantages, however, BPF is not well understood and lacks important theoretical background, such as a proof of convergence and insight into the resampling step based on interval analysis~\cite{Gning2013}. Also, its exact computational cost yet remains to be investigated and compared with traditional HF and PF.

\section{The Classical SIR Particle Filter}
\label{sec:sir} 
Recursive Bayesian importance sampling~\cite{Geweke1989} of an unobserved and discrete Markov process $\{\mathbf{x}_{k}\}_{k=1,\ldots ,K}$ is based on three components: (i) the measurement vector $\mathbf{Z}^k=\{\mathbf{z}_{1},\ldots ,\mathbf{z}_{k}\}$, (ii) the dynamics (i.e., state transition) probability distribution $p(\mathbf{x}_{k} | \mathbf{x}_{k-1})$, and (iii) the likelihood $p(\mathbf{z}_{k} | \mathbf{x}_{k})$. Then, the state posterior $p(\mathbf{x}_{k} | \mathbf{Z}^{k})$ at time $k$ is recursively computed as:
\begin{equation}
\underset{\text{posterior}}{\underbrace{p(\mathbf{x}_{k} | \mathbf{Z}_{k})}} = \frac{\overset{\text{likelihood}} {\overbrace{ p(\mathbf{z}_{k} | \mathbf{x}_{k})}}\,\, \overset{\text{prior}} {\overbrace{p(\mathbf{x}_{k} | \mathbf{Z}^{k-1})}}}{\underset{\text{normalization}}{\underbrace{p(\mathbf{z}_{k} | \mathbf{Z}^{k-1})}}}\, ,
\end{equation}
where the prior is defined as:
\begin{equation}
p(\mathbf{x}_{k} | \mathbf{Z}^{k-1}) = \int p(\mathbf{x}_{k} | \mathbf{x}_{k-1}) \, p(\mathbf{x}_{k-1} | \mathbf{Z}^{k-1}) \, \mathrm{d}\mathbf{x}_{k-1}.
\end{equation}
In the PF approach, the posterior at each time point $k$ is approximated by $N$ weighted samples (i.e., particles) $\{\mathbf{x}^i_k, w^i_k\}_{i=1,\ldots ,N}$. This approximation is achieved by drawing a set of particles from an importance function (i.e., proposal distribution) $\pi(\cdot)$ and updating their weights according to the dynamics PDF and the likelihood. This process is called sequential importance sampling (SIS)~\cite{Doucet2001}. However, SIS suffers from the \textit{weight degeneracy}, where small particle weights become even smaller and do not contribute to the posterior any more. To overcome this, a \textit{resampling} step is performed~\cite{Doucet2001} whenever the sample size falls below a preset threshold. Using the standard notation, as in Refs.~\cite{Bashi2003,Doucet2001}, the complete SIR algorithm is given in Algorithm~\ref{alg:sir}.

\begin{algorithm}
\caption{Sequential Importance Resampling (SIR)} \label{alg:sir}
\begin{algorithmic}[1]
\Procedure{SIR}{}
	\For{$i=1 \to N$} \Comment{Initialization, $k$=0} 
		\State $w_0^{i} \gets 1/N$
		\State Draw $\mathbf{x}_0^{i}$ from $\pi(\mathbf{x}_0)$
	\EndFor
	\For{$k=1 \to K$}  
		\For{$i=1 \to N$}  \Comment SIS step
			\State Draw a sample $\tilde{\mathbf{x}}_k^i$ from $\pi(\mathbf{x}_k | \mathbf{x}_{k-1}^i,\mathbf{Z}^{k})$
			\State Update the importance weights
			\State $\tilde{w}_k^i \gets w_{k-1}^i \frac{p(\mathbf{z}_{k} | \tilde{\mathbf{x}}_{k}^i)  p(\tilde{\mathbf{x}}_{k}^i | \mathbf{x}_{k-1}^i)}{\pi(\tilde{\mathbf{x}}_k^i | \mathbf{x}_{k-1}^i,\mathbf{Z}^{k})}$ 
		\EndFor
		\For{$i=1 \to N$} 
			\State $w_k^i \gets \tilde{w}_k^i / \sum_{j=1}^{N} \tilde{w}_k^j$
		\EndFor 
		
		\Comment Calculate the effective sample size
		\State $\widehat{N}_{\textrm{eff}} \gets 1 / \sum_{j=1}^{N} (w_{k}^{j})^2$ 
		\If{$\widehat{N}_{\textrm{eff}}<N_{\textrm{threshold}}$} \Comment Resampling step
				%\State Select index $s(i)\in \{1,...,N\}$ with probability $w^i_{k}$
				\State Sample a set of indices $\{s(i)\}_{i=1,\ldots ,N}$ distributed such that $\Pr[s(i)=l]=w_{k}^{l}$ for $l= 1 \to N$.
			\For{$i=1 \to N$}
				\State $\mathbf{x}_{k}^{i} \gets \tilde{\mathbf{x}}_{k}^{s(i)}$
				\State $w_{k}^{i} \gets 1/N$ \Comment Reset the weights
			\EndFor
		\EndIf
	\EndFor
\EndProcedure
\end{algorithmic}
\end{algorithm}

\section{The Piecewise Constant SIR Particle Filter}
\label{sec:asir} 
In classical SIR, all particle weights are updated according to the likelihood, which may impart a high computational load. Moreover, the computational cost scales linearly with the number of particles. Therefore, depending on the application, the likelihood evaluation often constitutes the most time-consuming part of a PF. 

To address this problem, we propose the pcSIR algorithm, which aims at reducing the computational cost of importance weight update by exploiting the nature of the particle function approximation underlying SIR~\cite{greengard1987fast}. We do this by grouping the particles into non-overlapping multi-dimensional \textit{bins} (i.e., Cartesian mesh cells in higher dimensions), which are then represented by only a single dummy particle positioned at the center of mass of the real particles in that bin. The center of mass is computed using the state vectors and weights of all particles within the bin and is solely used to \textit{represent} that bin by a single \textit{dummy particle}. This amounts to a first-order multipole expansion of the PDF approximated by the particles~\cite{greengard1987fast}. Higher-order approximations are easily possible by storing on the dummy particle not only the mean, but also higher-order moments of the particle distribution in the bin. However, the overall error of a PF is dominated by the Monte-Carlo sampling error, which is of order 1/2. A first-order function approximation is hence sufficient. 

The importance weight update is then only applied to the dummy particle. All other particles in the same bin are assigned the same weight that the dummy particle received. Thus, we approximate the likelihood by a mixture of uniform PDFs and bypass the costly likelihood update step for all particles. The pcSIR algorithm differs from SIR only in the SIS part, where the particles are binned and several averaging operations are performed. This makes it straightforward to implement pcSIR in any existing SIR code. The detailed pseudo-code is given in Algorithm~\ref{alg:asir}. 
\begin{algorithm}
\caption{Piecewise Constant Sequential Importance Resampling (pcSIR)} \label{alg:asir}
\begin{algorithmic}[1]
\Procedure{pcSIR}{}
	\For{$i=1 \to N$} \Comment{Initialization, k=0} 
		\State $w_0^{i} \gets 1/N$
		\State Draw $\mathbf{x}_0^{i}$ from $\pi(\mathbf{x}_0)$
	\EndFor
	\State Create $B$ bins of equal size $I_{1,\ldots ,B}$
	\For{$k=1 \to K$}  		
		\For{$i=1 \to N$} \Comment piecewise constant SIS step
			\State Draw a sample $\tilde{\mathbf{x}}_k^i$ from $\pi(\mathbf{x}_k | \mathbf{x}_{k-1}^i,\mathbf{Z}^{k})$
			 \State Assign $\tilde{\mathbf{x}}_k^i$  to a bin
		\EndFor
		\For{$j=1 \to B$} \Comment Visit all bins \\
		\Comment Create a \textit{representative} particle that has the mean values of the state vector of all particles in the same bin
		\State $\mathbf{x}_{\textrm{dum}} \gets \textrm{mean} 
		%\left (\tilde{\mathbf{x}}_k^{\left (1,...,N_{I_{j}} \right )}\right )
		\{\tilde{\mathbf{x}}_k^1,\ldots ,\tilde{\mathbf{x}}_k^{N_{I_{j}}}\}$
		\State Update the importance weights
			\State $w_{\textrm{dum}_k} \gets w_{\textrm{dum}_{k-1}} \frac{p(\mathbf{z}_{k} | \mathbf{x}_{\textrm{dum}})  p(\mathbf{x}_{\textrm{dum}} | \mathbf{x}_{k-1}^i)}{\pi(\mathbf{x}_{\textrm{dum}} | \mathbf{x}_{k-1}^i,\mathbf{Z}^{k})}$
		%\For{$i=1 \to N_{I_{j}}$}  			
		\For{all $\tilde{\mathbf{x}}_k^i$ in bin $I_j$}  			
			\State $w_k^i \gets w_{\textrm{dum}_k}$
		\EndFor
		\EndFor
		
		\Comment Calculate the effective sample size
		\State $\widehat{N}_{\textrm{eff}} \gets 1 / \sum_{j=1}^{N} (w_{k}^{j})^2$ 
		\If{$\widehat{N}_{\textrm{eff}}<N_{\textrm{threshold}}$} \Comment Resampling step
							\State Sample a set of indices $\{s(i)\}_{i=1,\ldots ,N}$ distributed such that $\Pr[s(i)=l]=w_{k}^{l}$ for $l= 1 \to N$.
			\For{$i=1 \to N$}
				\State $\mathbf{x}_{k}^{i} \gets \mathbf{x}_{k}^{s(i)}$
				\State $w_{k}^{i} \gets 1/N$ \Comment Reset the weights
			\EndFor
		\EndIf
	\EndFor
\EndProcedure
\end{algorithmic}
\end{algorithm}

The final function approximation used in pcSIR is related to BPF, where the box support is also approximated by a mixture of piecewise constant functions~\cite{Gning2010}. However, the theoretical motivation and the algorithmic implementation of this piecewise constant approximation is very different in pcSIR and in BPF. Gning \textit{et al.}~\cite{Gning2010,Gning2013} used interval analysis~\cite{Jaulin2001} to show that the uniform PDF approximation of the {\em posterior} becomes more accurate as the number of intervals increases. In pcSIR, the piecewise constant approximation of the {\em likelihood} is rooted in particle function approximation theory and can be understood as a first-order multipole expansion~\cite{greengard1987fast}. This dispenses with the need for interval analysis and provides a different algorithmic implementation and error analysis.

Unlike BPF, pcSIR still uses point particles. Thus, state estimation problems that result in narrow posterior densities can easily be handled by pcSIR, which is not the case for BPF. With pcSIR, we provide a simple way of using uniform PDFs to approximate the {\em likelihood} function, which eventually results in a satisfactory posterior representation through the Bayesian formulation. Moreover, it requires only few modifications to the classical SIR, which makes pcSIR an attractive choice for practical implementations. 

\subsection{Theoretical framework}

The pcSIR algorithm is a function-approximation algorithm. It divides the $n$-dimensional state space into $n$-dimensional bins. In each bin, a sufficiently differentiable likelihood function is approximated by a constant value. The error analysis of such piecewise constant approximations is well understood on the basis of Taylor's theorem for multivariate functions. 

For the sake of example, we present the theoretical framework of pcSIR with a focus on image processing. When processing a sequence of 2D images, the likelihood function $p(\mathbf{z}_{k}|\mathbf{x}_{k})$ is typically a two-dimensional function that is discretized over a finite set of particles. In SIR, the likelihood is approximated by $N$ particles, where the particle number $N$ defines the accuracy for the specific application. Therefore, the approximation error of SIR is denoted $\mathtt{E}_{SIR}(N)$. 
 
With pcSIR, in the considered application, only the positions of the particles play a role in the likelihood update. This allows pcSIR to bin the state space. Therefore, the approximation error in $p(\mathbf{z}_{k}|\mathbf{x}_{k})$ depends on both the number of particles $N$ and the maximum lengths of the bins $l_x$ and $l_y$ in both dimensions. Hence, the overall approximation error of pcSIR is denoted $\mathtt{E}_{pcSIR}(N,l_x,l_y)$. 
 
First, we analyze the effect of bin size on $\mathtt{E}_{pcSIR}(N,l_x,l_y)$. For that purpose, we consider two cases: The first considers bins of varying rectangular shapes (i.e., $l_x \neq l_y$). In this setting, we fix $N$ and let the approximation error depend on the bin lengths in both dimensions, hence $\mathtt{E}_{pcSIR}(l_x,l_y)$. In the second case, all cells are squares of edge length $l$. The pcSIR approximation error can then be expressed as $\mathtt{E}_{pcSIR}(l)$.

Second, we compare SIR with an pcSIR in which each bin corresponds to a single pixel in a ``pseudo''-tracking test case (see Section~\ref{sec:benchmark}) In this comparison, we assume Gaussian and uniform priors of different sizes. A smooth likelihood function is approximated by SIR and pcSIR and later applied to the prior. Thus, we obtain the estimation errors for the state. We call this experiment ``pseudo''-tracking, since by eliminating the explicit dynamics PDF, we can focus on the approximation error and its convergence with increasing $N$.

\subsection{The effect of cell size on $\mathtt{E}_{pcSIR}$}
The particle locations in a cell cannot be determined \textit{a priori} since the movement of the particles depends on the data. We hence assume that for small cells and statistically large numbers of particles, we have a uniform particle distribution within a cell. The approximation errors introduced by the pcSIR algorithm in 2D are described in detail in the Appendix. 

In pcSIR, the state space is decomposed into non-overlapping cells. Choosing an appropriate cell size is hence crucial for pcSIR. Similar to histogram filters~\cite{Thrun2005}, the accuracy of pcSIR is determined by the cell size. In the highest possible resolution, there is one particle per cell, which recovers the classical SIR algorithm. 

In image processing, it is convenient to choose the image pixels as the cells of pcSIR. This constitutes a good choice since in typical image-processing applications, the pixel size already reflects the sizes of the objects represented in the image in order not to under-sample the objects and not to store unnecessary data. We call pcSIR with single-pixel cells pcSIR-1x1. Due to the characteristics of the likelihood function, however, there may be cases where sub-pixel resolution or higher accuracy is needed. Therefore, we also investigate pcSIR-2x2, where each pixel is divided into four cells. In the following Section, we empirically benchmark the effect of cell size on pcSIR performance and accuracy. 

\section{Experimental Results}
\label{sec:benchmark}
We study the performance of pcSIR by considering a biological image-processing application: the tracking of sub-cellular (here, ``cell'' refers to the biological cell being imaged and is not to be confused with the pcSIR bin cells) objects imaged by fluorescence microscopy~\cite{akhmanova2005, komarova2009mammalian,helmuth2009deconvolving}. There, intracellular structures such as endosomes, vesicles, mitochondria, or viruses are labeled with fluorescent dyes and imaged over time with a confocal microscope. Many biological studies start from analyzing the dynamics of those structures and extracting parameters that characterize their behavior, such as average velocity, instantaneous velocity, spatial distribution~\cite{helmuth2010beyond,Shivanandan:2013}, motion correlations, etc. 

\subsection{Dynamics model}
The motion of sub-cellular objects can be represented by a variety of dynamics models, ranging from random walks to constant-velocity models to more complex dynamics where switching between motion types occurs~\cite{smal_media,godinez2012identifying}. 

Here, we use a nearly-constant-velocity model, which is frequently used in practice~\cite{smaltmi,rong2003survey}. The state vector in this case is $\mathbf{x}=(\hat{x}, \hat{y}, v_x, v_y, I_0)^T$, where $\hat{x}$ and $\hat{y}$ are the $x$- and $y$-positions of an object, $(v_x,v_y)$ its velocity vector, and $I_0$ its fluorescence intensity. 

\subsection{Likelihood / Appearance model}
Many sub-cellular objects are smaller than what can be resolved by the microscope, making them appear in a fluorescence image as diffraction-limited bright spots with an intensity profile given by the impulse-response function of the microscope, the so-called point-spread-function (PSF)~\cite{smaltmi,smal_media,helmuth2009deconvolving}. 

In practice, the PSF of a fluorescence microscope is well approximated by a 2D Gaussian~\cite{thomann2002automatic,zhang2007gaussian}. 
Object appearance in a 2D image is hence modeled as:
\begin{equation}
\label{eqn:object}
I(x,y;x_0, y_0) = I_0 \exp\left(-\frac{(x-x_0)^2 + (y-y_0)^2}{2\sigma^2_{\textrm{PSF}}}\right) + I_{\textrm{bg}},
\end{equation}     
where $(x_0, y_0)$ is the position of the object, $I_0$ is its intensity, $I_{\textrm{bg}}$ is the background intensity, and $\sigma_{\textrm{PSF}}$ is the standard deviation of the Gaussian PSF. Typical microscope setups yield images with pixel edge lengths corresponding to 60 to 200\,nm real-world length in the imaged sample. For the images used here, the pixel size is 67\,nm and the microscope has $\sigma_{\textrm{PSF}}=78$\,nm (or 1.16 pixels). During image acquisition, the ``ideal'' intensity profile $I(x,y)$ is corrupted by measurement noise, which in the case of fluorescence microscopy has mixed Gaussian-Poisson statistics. For the resulting noisy image $\mathbf{z}_k=Z_k(x,y)$ at time point $k$, the likelihood $p(\mathbf{z}_k|\mathbf{x}_{k})$ is:
\begin{equation} 
\label{eqn:likelihood}
p(\mathbf{z}_k|\mathbf{x}_{k}) \varpropto \exp\!\!\left(\!-\frac{1}{2\sigma^2_{\xi}}\!\!\sum_{(x_i, y_i)\in\mathbb{S}_{\mathbf{x}}}\!\!\!\!\!\!\left [Z_k(x_i, y_i)-I(x_i, y_i;\hat{x}, \hat{y})\right ]^2\!\!\right)\!\!,
\end{equation}     
where $\sigma_{\xi}$ controls the peakiness of the likelihood, $(x_i, y_i)$ are the integer coordinates of the pixels in the image, $(\hat{x}, \hat{y})$ are the spatial components of the state vector $\mathbf{x}_k$, and $\mathbb{S}_{\mathbf{x}}$ defines a small region in the image centered at the object location specified by the state vector $\mathbf{x}_k$. Here, $\mathbb{S}_{\mathbf{x}}=[\hat{x} - 3\sigma_{\textrm{PSF}}, \hat{x} + 3\sigma_{\textrm{PSF}}]\times[\hat{y} - 3\sigma_{\textrm{PSF}}, \hat{y} + 3\sigma_{\textrm{PSF}}]$.

\subsection{Experimental setup}
\begin{figure}[]
\centering
\includegraphics[width=0.8\columnwidth]{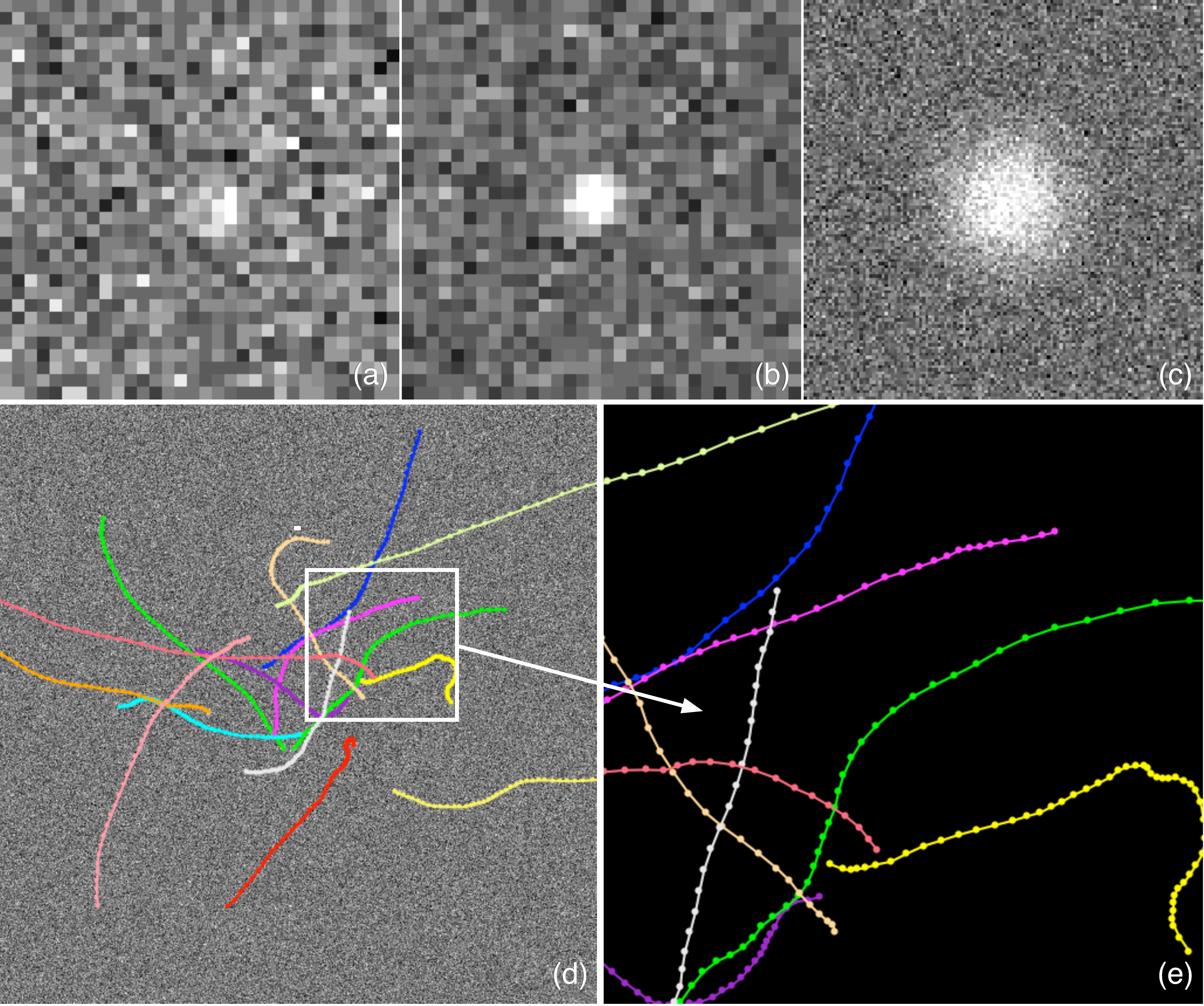}
\caption{Examples of object appearance for different object sizes and SNR: (a) $\sigma_{\textrm{PSF}} = 1.16$, SNR=2, (b) $\sigma_{\textrm{PSF}} = 1.16$, SNR=4, (c) $\sigma_{\textrm{PSF}} = 13$, SNR=2. (d/e): Typical object trajectories generated using the nearly-constant-velocity dynamics model.}
\label{image_data}
\end{figure}

We focus on single sub-cellular object tracking (a problem which is related to the ``track-before-detect'' problem~\cite{ristic2004beyond}) and compare pcSIR with SIR in two test cases, which differ in the size of the tracked object. We consider two different object sizes in order to compare cases where the likelihood is computationally cheap to evaluate with cases where this is more costly. 20 synthetic image sequences of different quality (i.e., signal-to-noise ratios, SNR) are generated by simulating a microscope. Each sequence is composed of 50 frames of size 512$\times$512 pixels. The movies show a single object moving according to the dynamics model. Examples are shown in Fig.~\ref{image_data}. 

The two object sizes correspond to $\sigma_{\textrm{PSF}}=1.16$ and $\sigma_{\textrm{PSF}}=13$, and are named ``small object tracking'' and ``large object tracking'', respectively (Fig.~\ref{image_data}(a-c)). The positions and directions of motion of the objects are randomly chosen within the image plane. The speed (i.e., the displacement in pixels per frame) is drawn uniformly at random over the interval $[2, 7]$ for large objects and over $[2, 4]$ for small objects. The SNR of the images of large objects is 2 (ca. 6 dB), that for small objects is 4 (ca. 12 dB). We use the SNR definition for Poisson noise~\cite{cheezum2001quantitative}. In the literature on sub-cellular object tracking, a SNR of 4 is considered critical, as for lower SNRs many of the available tracking methods fail~\cite{thomann2002automatic}.

\begin{figure}[]
\centering
\includegraphics[width=0.8\columnwidth]{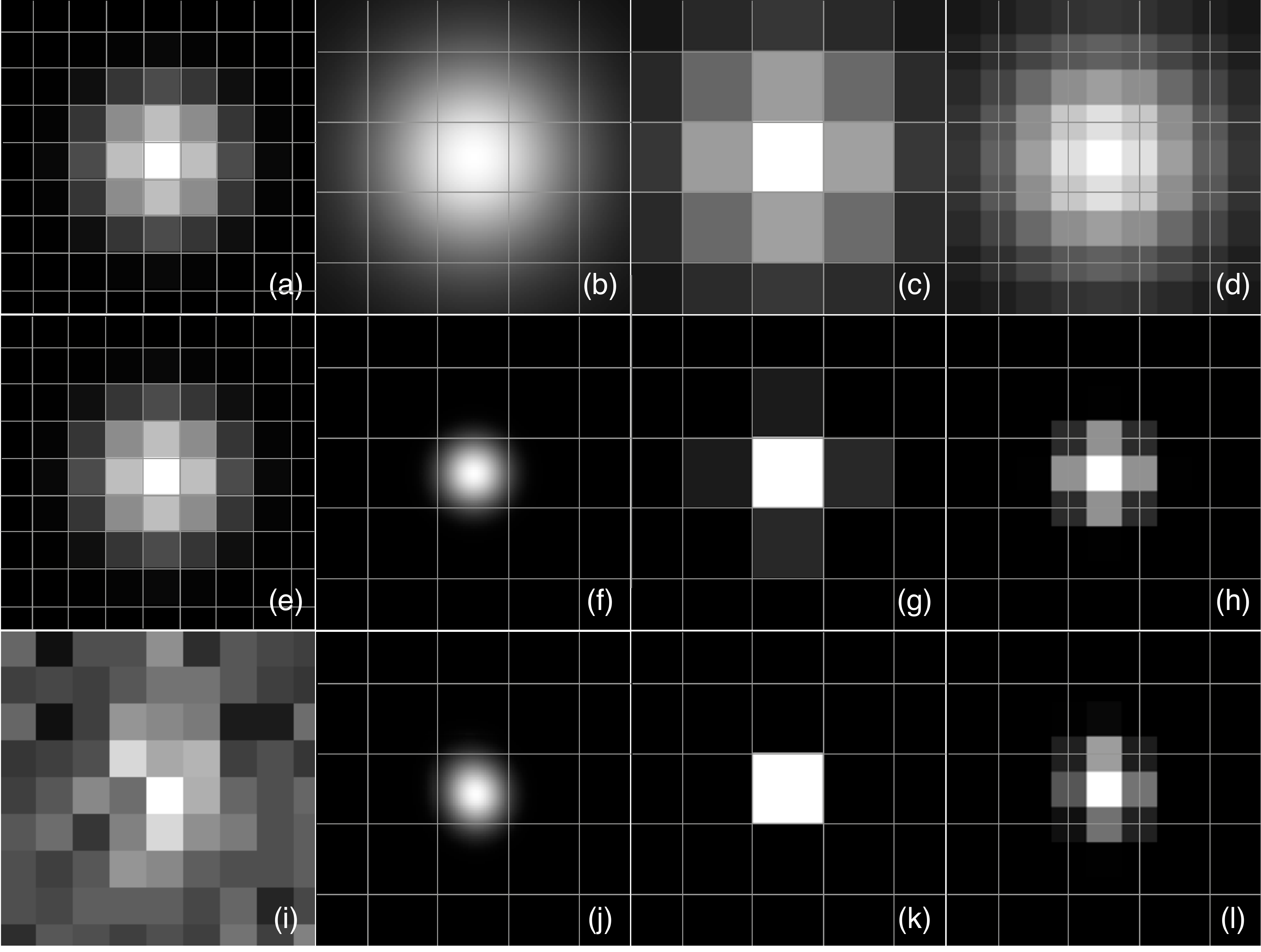}
\caption{Examples of likelihood profiles. The noise-free objects are shown in (a, e), and the noisy (SNR=2) object in (i) with $\sigma_{\textrm{PSF}} = 1.16$. We show the corresponding likelihood kernels (b, f, j), the approximated likelihoods used by pcSIR-1x1 (c, g, k), and the approximated likelihoods used by pcSIR-2x2 (d, h, l). In (b, c, d) the parameter $\sigma_{\xi}$ is 30, for the rest $\sigma_{\xi}=10$. The distance between the grid-lines corresponds to the size of the image pixel.}
\label{Likelihoods}
\end{figure}

Knowing the ground-truth object positions and those estimated by the PF, we quantify the tracking accuracy by the root-mean-square error (RMSE) in units of pixels. The likelihood kernel for the large objects has a support of 65$\times$65 pixels and is correspondingly costly to evaluate. The kernel for the small objects has a support of 9$\times$9 pixels and is cheaper to evaluate. Examples of noise-free and noisy object profiles, together with their likelihood kernels, are shown in Fig.~\ref{Likelihoods}.

Using double-precision arithmetics, a single PF particle requires 52\,KB (i.e., six doubles and one integer) of computer memory. The particles are initialized at the ground-truth location and all tests are repeated 50 times for different realizations of the image-noise process on a single core of a 12-core Intel\textregistered \,  Xeon\textregistered \, E5-2640 2.5\,GHz CPU with 128\,GB DDR3 800\,MHz memory on MPI-CBG's MadMax computer cluster. All algorithms are implemented in Java (v. 1.7.0\_13) within the Parallel Particle Filtering (\texttt{PPF}) library~\cite{demirel2013ppf}. The results are summarized in Figs.~\ref{asirvssir_bigObj} and \ref{asirvssir_smallObj} for large and small objects, respectively.

\subsection{Results}
\begin{figure}[]
\centering
\includegraphics[width=0.92\columnwidth]{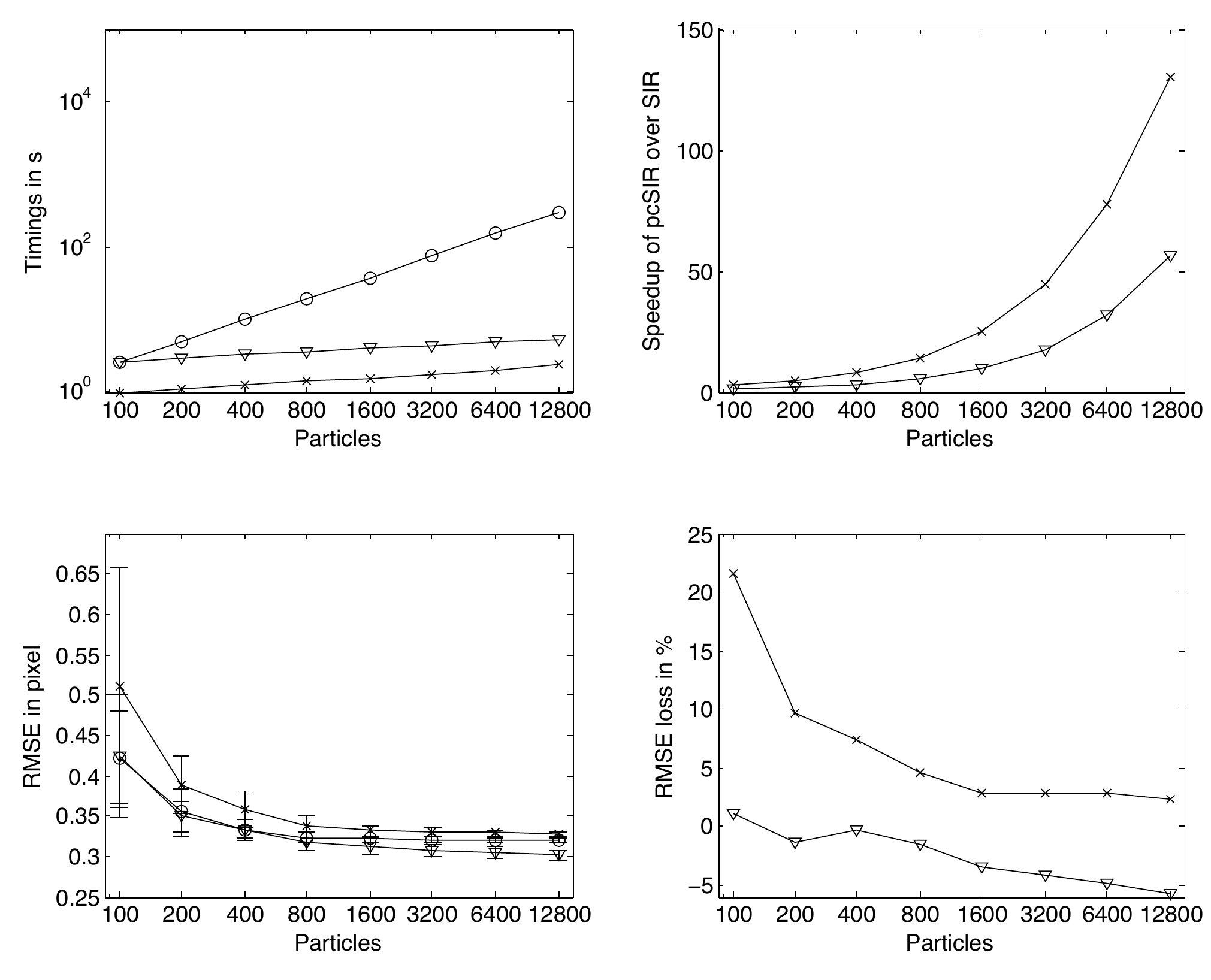}
\caption{Runtime performance and tracking accuracy of pcSIR-1x1 ($\times$) and pcSIR-2x2 ($\triangledown$) compared with SIR ($\circ$) for a 65 pixel wide likelihood kernel. The number of particles used starts from 100 and is doubled for each case until 12\,800. The timings of all three methods are presented in $\log$-$\log$ scale (upper left), whereas the relative speedups of the pcSIR methods over SIR are shown in the upper-right plot. The accuracy loss (lower right) of pcSIR-1x1 drops rapidly as the number of particles in the system is increased. Error bars show standard deviations across the 50 repetitions of each experiment.}
\label{asirvssir_bigObj}
\end{figure}

When tracking large objects (Fig.~\ref{asirvssir_bigObj}), both pcSIR versions provide significant speedups over the classical SIR algorithm. For 12\,800 particles, pcSIR-1x1 is more than two orders of magnitude faster than SIR with a 2.4\% loss in tracking accuracy. pcSIR-2x2 provides an up to 5.8\%
better tracking accuracy than SIR while running over 50 times faster. Since SIR is also an approximation of the actual posterior distribution, in some cases pcSIR may provide a better representation of the posterior and thus a higher tracking accuracy. This phenomenon has been previously described~\cite{koblentspopulation}. 

\begin{figure}[]
\centering
\includegraphics[width=0.92\columnwidth]{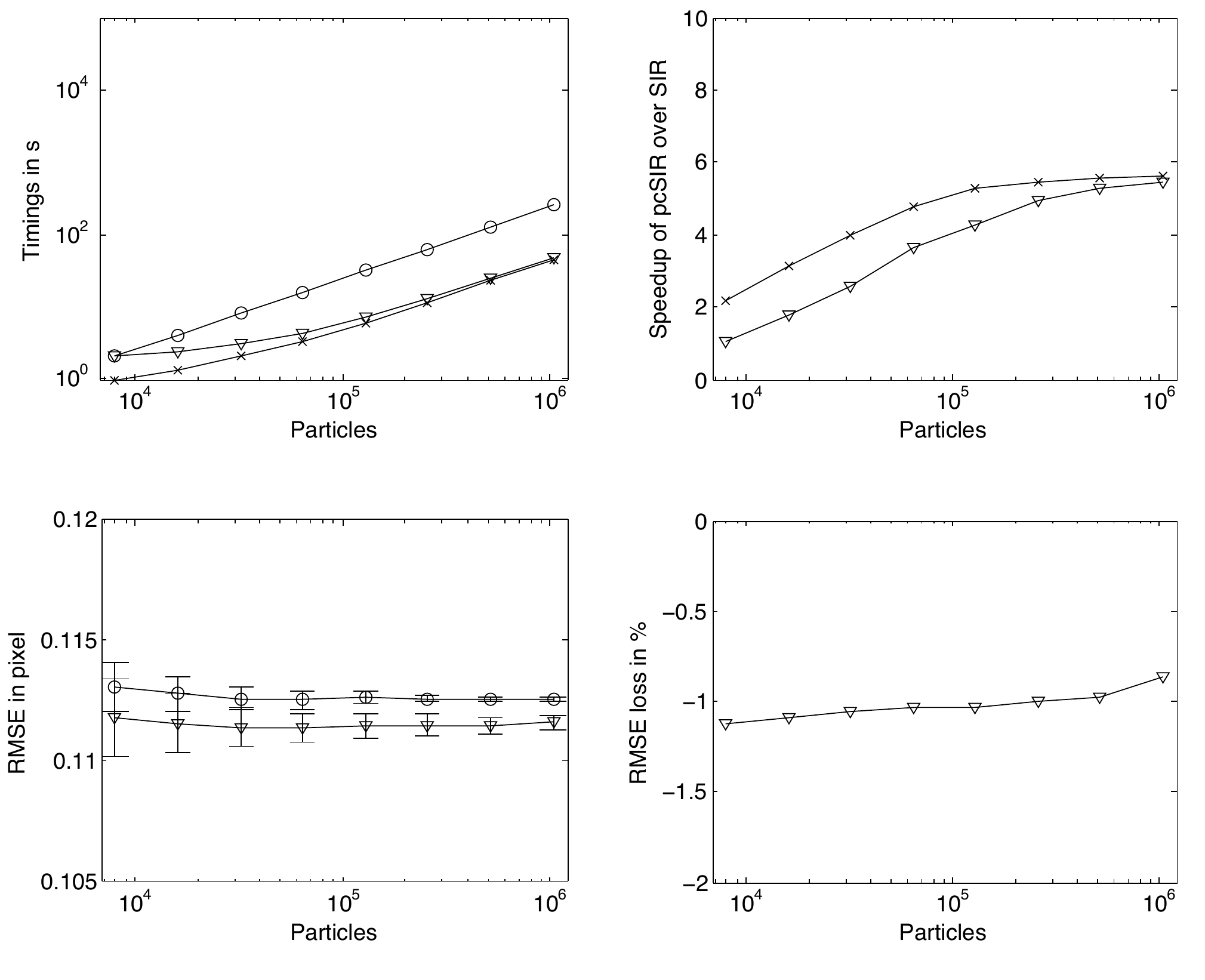}
\caption{Runtime performance and tracking accuracy of pcSIR-1x1 ($\times$) and pcSIR-2x2 ($\triangledown$) compared with SIR ($\circ$) for a nine-pixel wide likelihood kernel. The number of particles used starts from 8\,000 and is doubled for each case until 1\,024\,000. The timings of all three methods are presented in $\log$-$\log$ scale (upper left), whereas the relative speedups of the pcSIR methods over SIR are shown in the upper-right plot. For the accuracy comparisons (lower left), we show only the results for pcSIR-2x2 and SIR, since pcSIR-1x1's coarse bin resolution results in a 150\% worse tracking accuracy than SIR. Error bars show standard deviations across the 50 repetitions of each experiment.}
\label{asirvssir_smallObj}
\end{figure}

When tracking small objects, the likelihood support requires sub-pixel resolution and the effect of bin size is more visible (Fig.~\ref{asirvssir_smallObj}).  pcSIR-1x1 uses rather coarse bins compared to the likelihood support (Fig.~\ref{Likelihoods}), resulting in a pronounced loss of tracking accuracy. Visually, however, the trajectories produced by SIR and pcSIR-1x1 are virtually indistinguishable, since the tracking accuracy of pcSIR-1x1 is still in the sub-pixel regime (about 0.27 pixel). When finer bins (pcSIR-2x2) are used, the tracking accuracy of pcSIR is again better than that of SIR, and pcSIR runs more than five times faster than SIR.   
 
\subsection{Convergence of SIR and pcSIR}
\begin{figure}[]
\centering
\includegraphics[width=\columnwidth]{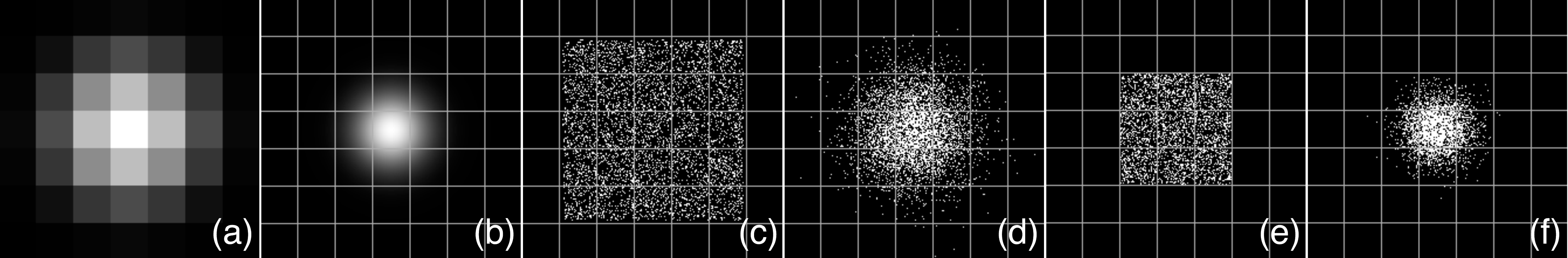}
\caption{The ``pseudo''-tracking experiment: (a) the object with $\sigma_{\textrm{PSF}} = 1.16$, SNR=2; (b) the corresponding likelihood with $\sigma_{\xi} = 20$; (c) a uniform prior of support 5$\times$5 pixel; (d) a Gaussian prior with $\sigma_{\textrm{prior}}=0.5$; (e) a uniform prior of support 3$\times$3 pixel; (f) a Gaussian prior with $\sigma_{\textrm{prior}}=0.8$. Thin white lines indicate the image pixel grid.}
\label{fig:spotandlikelihood}
\end{figure}

Both SIR and pcSIR employ particle approximations of a smooth, differentiable function, the order of accuracy of which depends on the number of  particles $N$.  
In order to eliminate uncertainties resulting from the dynamics model, we assume the prior $p(\mathbf{x}_k|\mathbf{Z}^{k-1})$ to be either a uniform distribution over 3$\times$3 or 5$\times$5 pixels, or a Gaussian with $\sigma_{\textrm{prior}}=\{0.5,0.8\}$, respectively. We then evaluate the likelihood  in Eq.~(\ref{eqn:likelihood}) with $\sigma_{\xi} =20$ using both SIR and pcSIR. We call this a ``pseudo''-tracking experiment.
%The continuous likelihood function is defined as:
%\begin{equation}
%\mathbb{L}(x,y) = \frac{1}{2 \pi \sigma^2_{\xi}} \exp \left (-\frac{x^2+y^2}{2 %\sigma^2_{\xi}}\right),
%\end{equation}
The object is a single PSF (Eq.~(\ref{eqn:object})) with $\sigma_{\textrm{PSF}}=1.16$. Visualizations of the object, likelihood, and prior are shown in Fig.~\ref{fig:spotandlikelihood}.

\begin{figure}[]
\centering
\includegraphics[width=0.44\columnwidth]{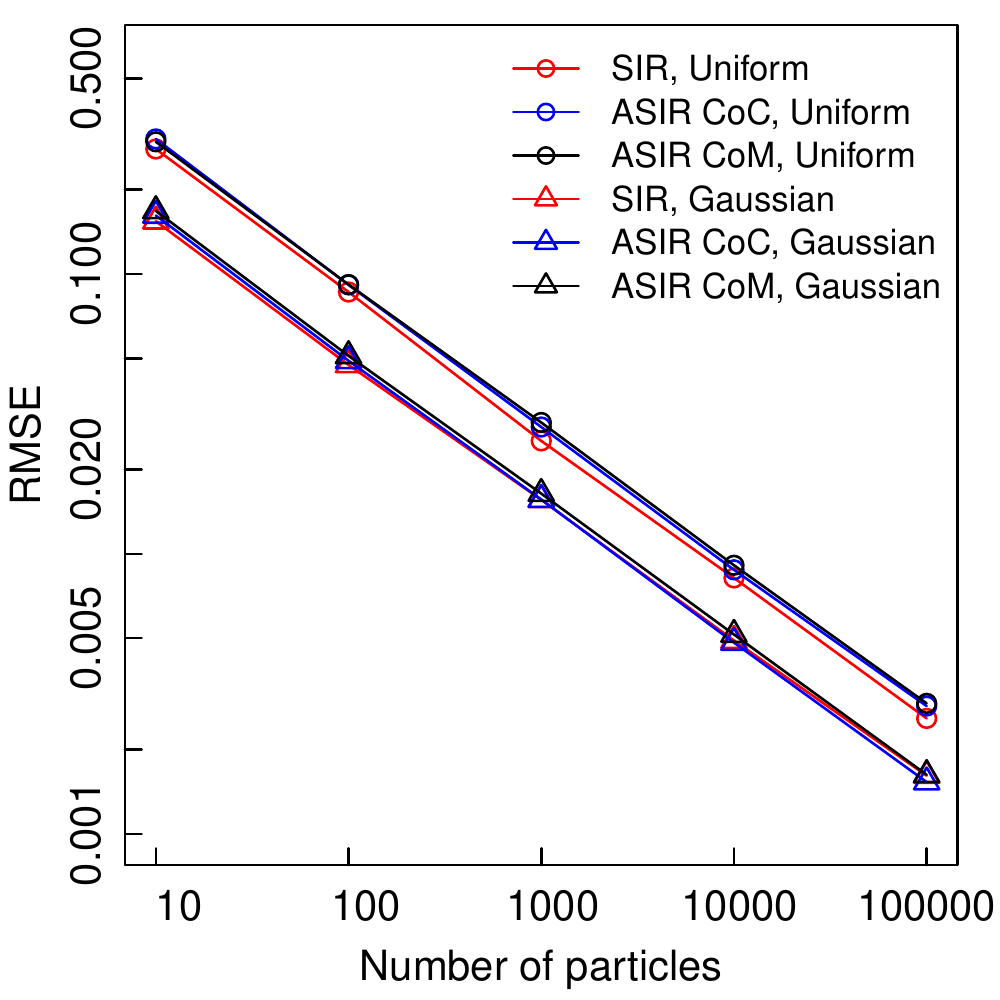}
\includegraphics[width=0.44\columnwidth]{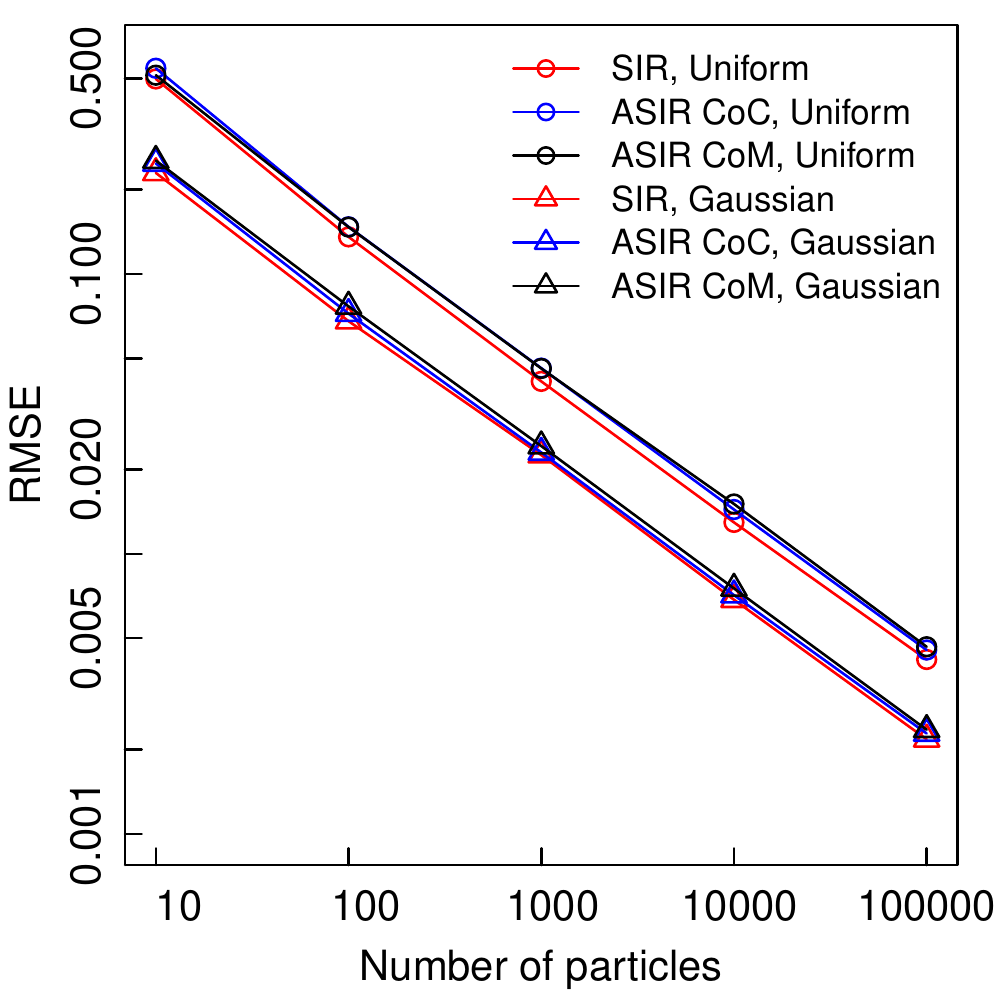}
\caption{The ``pseudo''-tracking experiment results for the Gaussian prior with $\sigma_{\textrm{prior}}=0.5$ and the uniform prior with 3$\times$3-pixel support (left), and for the Gaussian prior with $\sigma_{\textrm{prior}}=0.8$ and the uniform prior with 5$\times$5-pixel support (right). The state estimation errors of pcSIR relative to SIR range between $-12\%\ldots +6\%$. The difference between pcSIR and SIR decreases as $N$ increases. Both pcSIR and SIR converge with increasing number of particles. The RMSE error is reduced by about 30\% every time the number of particles doubles, corresponding to a convergence order of $\sqrt{N}$, as expected for a Monte Carlo method. Error bars are below symbol size.}
\label{fig:asirconvergence05}
\end{figure}

We compare two versions of pcSIR, which differ in the placement of the dummy particles: In pcSIR-CoC, the dummy particles are placed at the geometric centers of the bins, whereas in pcSIR-CoM, the centers of mass of the state vectors of all particles inside that bin are used. Each convergence experiment is repeated 1000 times for different realizations of the random process, and the number of particles is increased up to 100\,000. We quantify the RMSE of the state estimation as a function of the number of particles used. The resulting convergence plots for pcSIR and SIR are shown in Fig.~\ref{fig:asirconvergence05}.
% and \ref{fig:asirconvergence08}.

We observe no significant differences between SIR and the two pcSIR variants. The error of pcSIR-CoM is always slightly lower than that of pcSIR-CoC. SIR is generally the most accurate, but is outperformed by pcSIR-CoM in some cases, confirming our experimental tests as well as the findings in Ref.~\cite{koblentspopulation}. As $N$ increases, the errors of all methods decrease at the same rate. In all cases, however, the runtimes of both pcSIR variants were significantly less than that of SIR.
 
\section{Conclusions}
\label{sec:conclusion}
We proposed a fast approximate SIR algorithm, called pcSIR. pcSIR is based on spatially binning particles in \textit{cells} and representing each cell by a single dummy particle at the center of mass of the cell's particle distribution, carrying the average state vector of all particles in that cell. This approximates the likelihood by a first-order multipole expansion~\cite{greengard1987fast}. pcSIR significantly reduces the computational cost of SIR and enables tackling larger problems as well as tackling mid-size problems in real time. In some configurations, especially when sub-pixel resolution is used for the bins, pcSIR may yield more accurate results than SIR. 

We performed both theoretical and experimental error analysis of pcSIR. We showed that the error in the posterior decreases as the number of particles increases. Moreover, pcSIR converges at the same rate as SIR, since the Monte-Carlo sampling error masks the error from the function approximation. We presented theoretical upper bounds on the likelihood approximation error as a function of cell size in pcSIR. 

We experimentally tested the tracking accuracy and runtime performance of two pcSIR variants for image processing: pcSIR-1x1 and pcSIR-2x2. In our benchmarks, pcSIR showed significant speedups over SIR. As more particles are used, the relative speedup over SIR seems to grow exponentially for large-object tracking scenarios, where the likelihood is costly to evaluate. In the presented benchmarks with 12\,800 particles, SIR required 5 minutes to track the large object through a 50-frame 2D image sequence. pcSIR-1x1 needed only 2.3 seconds to accomplish the same task at the expense of a 2.4\% smaller accuracy. pcSIR-2x2 completed the task in 5.3 seconds with a 5.8\% better tracking accuracy than SIR. This improvement stems from the fact that for some posterior distributions, the piecewise constant likelihood approximate of pcSIR may be a more regular representation than that generated by SIR. This is a known phenomenon~\cite{koblentspopulation}. The relative speedups of the two pcSIR variants over classical SIR were 130-fold and 57-fold for 12\,800 particles, respectively. For larger numbers of particles, we expect even larger speedups. 

For small-object tracking, both pcSIR variants showed an average 5-fold improvement in execution time for the largest tested particle number. However, the tracking accuracy of pcSIR-1x1 is greatly reduced, since the likelihood function has a narrow support that is not well sampled by the coarse bins. While the errors are in the range of 150\%, they are barely visible in the final trajectories since the average RMSE is only about 0.27 pixels. Interestingly, pcSIR-2x2 shows improvements both in overall runtime (5-fold) and in tracking accuracy (1\%), which suggests that pcSIR-2x2 may be a good algorithm for tracking small objects. 

We believe that pcSIR can be used in many PF applications that require large numbers of particles, costly likelihood evaluations, or real-time performance. When tracking accuracy is not critical, pcSIR-1x1 can offer orders of magnitude speedup in image-processing applications. If a loss in tracking accuracy is undesired, pcSIR-2x2 still offers significant speedups while in some cases even improving accuracy over SIR. In other applications, one can adjust the size of the averaging bins according to the desired accuracy. Future improvements could involve adaptive bin sizes. 

\section*{Appendix: Approximation error of pcSIR in 2D}
\label{app:asir1}
Approximating integrable functions by piecewise constant functions is well understood in mathematics on the basis of Riemann integral theory~\cite{Davis1967,Thomas1984}. We formulate the approximation error $\mathtt{E}_{pcSIR}(l_x,l_y)$ of pcSIR with rectangular cells and then simplify it to $\mathtt{E}_{pcSIR}(l)$ for square cells. All results can be extended to higher-dimensional settings. 
%\begin{mytheorem} 
%\label{theorem:asir1}

Let the likelihood $p(\mathbf{z}_{k}|\mathbf{x}_{k})$ be a twice continuously differentiable function $f(x,y)$ within a domain $D \in \mathbb{R}^{[x_0,x_{n}]\times[y_0,y_{m}]}$, which is divided into $B=n\times m$ non-overlapping rectangular cells. Further, $l_{k_i}$ and $l_{k_j}$ denote the width (i.e., in $x$-direction) and the height (i.e., in $y$-direction) of cell $I_k$ in $D$, where $D= \bigcup_{k=1}^{B}I_k$. The indices $i$ and $j$ are given by $i=1,\ldots ,n$ and $j=1,\ldots ,m$, and the maximum side lengths in both dimensions are defined as $l_x = \max_{k_i}(l_{k_i})$ and $l_y = \max_{k_j}(l_{k_j})$ where $l_{k_i}=x_k-x_{k-1}$ and $l_{k_j}=y_k-y_{k-1}$. Then, the total approximation error $\mathtt{E}_{pcSIR}(l_x,l_y)$ of the likelihood in $D$ obtained by pcSIR (Algorithm~\ref{alg:asir}) is bounded by:
\begin{equation*}
\begin{split}
	\mathtt{E}_{pcSIR}(l_x,l_y)  \leq & \frac{1}{24} \left [  \max_{[D]}|f_{xx}| \, l_x^3 \, l_y + \max_{[D]}|f_{yy}| \, l_x l_y^3   \right ] ,
\end{split}
\end{equation*}
where $\max_{[x_0,x_{n}]}|f_{xx}|$ and $\max_{[y_0,y_{m}]}|f_{yy}|$ are the maxima of the absolute values of $\frac{\partial^2 f}{\partial x^2}$ and $\frac{\partial^2 f}{\partial y^2}$ in $D$, respectively. 

This result can be derived by mid-point Riemann-sum approximation of an integral. While the dummy particle does not have to be located at the center of a cell, for the sake of simplicity of the derivation, we assume that pcSIR uses the mid-point for piecewise constant likelihood approximation. Assume that $f(x,y)$ is twice continuously differentiable in region $D \in \mathbb{R}^{[x_0,x_{n}]\times[y_0,y_{m}]}$ and the following partial derivatives are defined: $\frac{\partial^2 f}{\partial x^2}=f_{xx}$, $\frac{\partial^2 f}{\partial y^2}=f_{yy}$ and $\frac{\partial^2 f}{\partial x \partial y}=f_{xy}$. The approximation error $\mathtt{E}_{I_{k}}(l_x,l_y)$ can be calculated by integrating the multivariate Taylor approximation
\begin{equation}
\label{eqn:asir1}
\begin{split}
	\mathtt{E}_{I_{k}}(l_x,l_y) &=  f(x,y)-f(a,b) \\
	& =  f_x(a,b)(x-a)+f_y(a,b)(y-b) \\
	& + \frac{1}{2!} [f_{xx}(a,b)(x-a)^2 + f_{yy}(a,b)(y-b)^2 \\
	& + 2 f_{xy}(a,b)(x-a)(y-b)]
\end{split}
\end{equation}
over the two-dimensional interval $I_k=[x_{k-1},x_k]\times[y_{k-1},y_k]$, where $a=\frac{x_{k-1}+x_k}{2}$, $b=\frac{y_{k-1}+y_k}{2}$, and $D= \bigcup_{k=1}^{B}I_k$, hence:
\begin{equation}
\label{eqn:asir1}
\begin{split}
	\mathtt{E}_{I_{k}}(B) =&  f_x(a,b) \int \!\!\! \int_{I_k} (x-a) \,\mathrm{d}x\,\mathrm{d}y \\ 
				     + & f_y(a,b)\int \!\!\! \int_{I_k} (y-b) \,\mathrm{d}x\,\mathrm{d}y \\
	+ & \frac{1}{2!} \left [ f_{xx}(a,b)  \int \!\!\! \int_{I_k} (x-a)^2 \,\mathrm{d}x\,\mathrm{d}y \right. \\
	+ & f_{yy}(a,b)  \int \!\!\! \int_{I_k} (y-b)^2 \,\mathrm{d}x\,\mathrm{d}y \\
         + & \left. 2 f_{xy}(a,b)  \int \!\!\! \int_{I_k}  (x-a)(y-b) \,\mathrm{d}x\,\mathrm{d}y  \right ].
\end{split}
\end{equation}
Substituting $a$ and $b$ in Eq.~(\ref{eqn:asir1}), we find: 
\begin{equation}
\label{eqn:asir2}
\begin{split}
	\mathtt{E}_{I_{k}}(l_x,l_y) = & \frac{1}{2!} \left [ f_{xx}(a,b)  \int \!\!\! \int_{I_k} \left (x-\frac{x_{k-1}+x_k}{2} \right)^2 \,\mathrm{d}x\,\mathrm{d}y \right. \\
	+ & \left. f_{yy}(a,b)  \int \!\!\! \int_{I_k} \left (y-\frac{y_{k-1}+y_k}{2}\right)^2 \,\mathrm{d}x\,\mathrm{d}y \right ].
\end{split}
\end{equation}
We substitute $l_{k_i}=x_k-x_{k-1}$, $l_{k_j}=y_k-y_{k-1}$ and evaluate the integrals. Then Eq.~(\ref{eqn:asir2}) becomes
\begin{equation}
\label{eqn:asir3}
\begin{split}
	\mathtt{E}_{I_{k}}(l_x,l_y) =  \frac{1}{24} \left [ f_{xx}(a,b)\, l_{k_i}^3 \, l_{k_j} + f_{yy}(a,b)\, l_{k_j}^3 \, l_{k_i} \right ].
\end{split}
\end{equation}
Next, we sum the absolute values of the partial errors in all $I_k$ regions in order to provide an upper bound on the total approximation error in the closed region $[D]$ as:
\begin{equation}
\label{eqn:asir4}
\begin{split}
	\mathtt{E}(l_x,l_y)  \leq & \frac{1}{24} \left [  \max_{[D]}|f_{xx}| \, \max_{k_i}(l_{k_i})^3 \,\max_{k_j}(l_{k_j}) \right. \\
	 &\quad \left.+ \max_{[D]}|f_{yy}| \, \max_{k_j}(l_{k_j})^3 \, \max_{k_i}(l_{k_i})  \right ]. 
\end{split}
\end{equation}
By substituting $l_x$ and $l_y$ into Eq.~(\ref{eqn:asir4}), we find the total error in the closed region $[D]$:
\begin{equation}
\label{eqn:asir4.5}
\begin{split}
	\mathtt{E}(l_x,l_y)  \leq & \frac{1}{24} \left [  \max_{[D]}|f_{xx}| \, l_x^3 \, l_y + \max_{[D]}|f_{yy}| \, l_x l_y^3   \right ]. 
\end{split}
\end{equation}

%\end{myproof}
%
%\section{Proof of the Extension of Theorem~\ref{theorem:asir1}}
%\label{app:asir2}
For equi-sized square cells, a tighter bound for the approximation error can be derived by repeating the steps that lead to Eq.~(\ref{eqn:asir3}). The derivation diverges here by taking the minimum possible value for the side lengths $l_{k_i}$ and $l_{k_j}$ of the small interval $I_k$ in region $D \in \mathbb{R}^{[x_0,x_{n}]\times[y_0,y_{m}]}$, where $D= \bigcup_{k=1}^{B}I_k$. When $I_k$ is a square with $l=l_{k_i}=x_k-x_{k-1}=l_{k_j}=y_k-y_{k-1}$, we obtain the bound on $\mathtt{E}_{I_{k}}(l)$ as:
\begin{equation} 
\label{eqn:asir5}
\begin{split}
	\mathtt{E}_{l_k}(l)  \leq & \frac{l^4}{24} \left [  f_{xx}  + f_{yy}  \right ]. 
\end{split}
\end{equation}
By summing the absolute values of the errors in all $I_k$ regions, we get the total error in closed region $[D]$:
\begin{equation} 
\label{eqn:asir6}
\begin{split}
	\mathtt{E}(l)  \leq & \frac{l^4}{24} \left [  \max_{[D]}|f_{xx}|  + \max_{[D]}|f_{yy}|  \right ].
\end{split}
\end{equation}

\section*{Acknowledgments}
The authors thank the MOSAIC Group (MPI-CBG, Dresden) for fruitful discussions and the MadMax cluster team (MPI-CBG, Dresden) for operational support. \"{O}mer Demirel was funded by grant \#200021--132064 from the Swiss National Science Foundation (SNSF), awarded to I.F.S. Ihor Smal was funded by a VENI grant (\#639.021.128) from the Netherlands Organization for Scientific Research (NWO).

\bibliographystyle{unsrt}
\bibliography{particle_filtering}

\end{document}